# Understanding the Essential Nature of the Hydrated Excess Proton Through Simulation and Interpretation of Recent Spectroscopic Experiments

Paul B. Calio, Chenghan Li, and Gregory A. Voth*

Department of Chemistry, Chicago Center for Theoretical Chemistry, James Franck Institute, and Institute for Biophysical Dynamics, The University of Chicago, 5735 South Ellis Avenue, Chicago, Illinois 60637, United States

Abstract

Two-dimensional infrared spectroscopy experiments have presented new results regarding the dynamics of the hydrated excess proton (aka "hydronium" cation solvated in water). It has been suggested by these experiments that the hydrated excess proton has an anisotropy reorientation timescale of 2.5 ps, which can be viewed as being somewhat long lived. Through the use of both the reactive molecular dynamics Multistate-Empirical Valence Bond method and Experiment Directed Simulation *Ab Initio* Molecular Dynamics we show that timescales of the same magnitude are obtained that correspond to proton transport, while also involving structural reorientations of the hydrated proton structure that correspond to the so-called "special pair dance". The latter is a process predicted by prior computational studies in which the central hydrated hydronium in a distorted Eigen cation ($H_9O_4^+$) structure continually switches special pair partners with its strongly hydrogen-bonded neighboring water molecules. These dynamics are further characterized through the time-evolution of instantaneous normal modes. It is concluded that the hydrated excess proton has a spectral signature unique from the other protons in the hydrated proton complex. However, the results call into question the use of a static picture based on a simple effective one dimensional potential well to describe the hydrated excess proton in water. Instead, they more conclusively point to a distorted and dynamic Eigen cation as the most prevalent hydrated proton species in acid solutions of dilute to moderate concentrations.

Introduction

The hydrated excess proton (aka "hydronium cation" plus nearby solvating water molecules) is pervasive in systems relevant to complex problems, e.g., proteins[1-4] and renewable energy materials.[5-7] Characterizing its essential solvation and transport behavior has been an active research area for more than two centuries. In bulk water, the hydrated excess proton has an anomalously high diffusion coefficient in comparison to other +1 cations,[3, 8] which is typically described via the Grotthuss mechanism,[9-10] where the net positive charge defect associated with an excess proton translocates over large distances through the breaking and forming of covalent bonds. In addition, the solvation structure of hydrated excess proton is sometimes described by the limiting cases of either an Eigen cation[11] (a hydronium ion, $H_3O^+$, core with strong hydrogen bonds to its three surrounding water molecules) or less often as a Zundel cation[12] (an excess proton equally solvated by two flanking water molecules). However, the exact nature of the hydrated excess proton hopping mechanism and most stable solvation structures represents an area of ongoing research interest (see, e.g., refs [13-14]).

The solvation and transport properties of hydrated excess protons have been the topic of numerous simulation studies, see, e.g., refs.[15-20] The dominant proton transfer (PT) mechanism is believed to be an Eigen-Zundel-Eigen (EZE) mechanism, where a "distorted" Eigen cation is the most stable structure and the Zundel cation is mainly an intermediate complex.[21-22] In this vein, proton transport occurs through the cleavage of a hydrogen bond in the second solvation shell, suggesting that Grotthuss hopping occurs via a step-wise process. Although the EZE proton hopping process represents the predominant mechanism, the simulation literature does include suggestions of Zundel-Zundel conversions,[15, 18, 23-24] although in some of these cases the underlying model appears to introduce a bias toward such a picture.

In actuality, however, the solvation structure of the hydrated excess proton is more complex than either Eigen or Zundel cations. In fact, the hydrated excess proton is seen in a broad range of configurations, making Eigen or Zundel cations only limiting structures and very difficult to deconvolute. For example, Tuckerman et al.[15, 25] utilized *ab initio* molecular dynamics (AIMD) simulations of an excess proton in 32 waters to confirm the presence of a "special pair" between the hydronium and nearby water molecules, which is very characteristic of a Zundel cation. Later molecular dynamics (MD) studies characterized the structure of the solvated proton as a distorted Eigen cation, where the three-fold symmetry is broken due to the distortion of the special pair.[17, 22, 26] In particular, classical multistate empirical valence bond (MS-EVB) and AIMD simulations were used to confirm that the identity of the special pair is not static, but instead switches with the other two hydrating water molecules in the Eigen cation on timescales of tens of femtoseconds. This "special pair dance" is a process in which the central hydrated hydronium in an Eigen cation ($H_9O_4^+$) structure continually switches special pair partners with its strongly hydrogen-bonded neighboring water molecules on a timescale of tens of femtoseconds. Hence, these studies suggested that the primary structure of the hydrated proton in water is, on average, a "distorted" Eigen cation, but also that this structure is quite dynamic among the three possible special pairs.

Elucidating the solvation structure of the hydrated excess proton is further complicated by the difficulty in correlating experimental infrared frequencies with structural information obtained from MD simulations. Although gas-phase results have been informative in correlating the two,[27-] [30] the condensed phase introduces further complications due to thermal and quantum fluctuations.[17, 31] When compared to the pure water absorption spectrum, four notable features can be seen in the acidic IR spectrum:[32-34] (1) a red shift in the O-H peak of bulk water correlating to stronger hydrogen bonding environments due to the excess proton; (2) an acid continuum from 2000-3200 cm$^{-1}$, which is most recently ascribed to more distorted Eigen-like configurations; (3) a peak at 1200 cm$^{-1}$ corresponding the proton transfer mode (PTM) between two flanking waters; and (4) a peak at 1750 cm$^{-1}$ corresponding to flanking water bend.

In the past few years, non-linear spectroscopy experiments have been pioneering efforts to understand the hydrated excess proton.[35-40] For example, experimental studies of acid clusters in acetonitrile mixtures[36, 40] observed the PTM $|1\rangle \rightarrow |2\rangle$ transition was greater than the $|0\rangle \rightarrow |1\rangle$ transition. This result was used to propose a 1-dimensional potential energy surface (PES) picture for the PTM that had a symmetric double-well structure, a possible characteristic of a Zundel cation. In water-acid solution,[35] Tokmakoff and co-workers used two-dimensional infrared spectroscopy to excite the O-H stretching vibrations around 3150 cm$^{-1}$ and detected spectral responses within a spectrum ranging from 1500 – 4000 cm$^{-1}$. By assigning 1750 cm$^{-1}$ to the bending vibration of the flanking waters of the Zundel complex, they suggested that the population of the Zundel cation is larger than previously proposed in theoretical studies, leading them to conclude that it may serve as more than a simple PT intermediate. Their subsequent studies[37] using 2D IR further (and importantly to the present topic) suggested that the effective potential energy surface (PES) is not a symmetric double-energy well characteristic of a symmetric Zundel cation, but rather incorporates a distortion into the underlying PES of the hydrated excess proton (and these authors called it a "distorted Zundel" structure). Additionally, by examining data obtained from parallel and perpendicular 2D IR spectra at 1750 cm$^{-1}$, an anisotropy timescale of ~2.5 ps was observed for 2M HCl solutions,[38] most likely correlating with irreversible proton transport.[41] In defining the excess proton as being "Zundel-like" in these studies, these researchers emphasized that the excess proton was shared between two flanking waters, synonymously known as a special pair.

Experimental IR spectra of acidic solutions, however, reveal time-averaged structures. Moreover, it is not uncommon for simulation studies to use static configurations to correlate vibrational motion with infrared frequency.[32-34, 42-45] Also, when using static configurations, simulation studies commonly implement a local $\delta$ parameter to characterize the solvation structure of the hydrated proton. This $\delta$ parameter conveys information on how the excess proton is shared between its flanking water molecules in the special pair and is defined as

$$\delta = |r_{O^*H} - r_{O^{SP}H}| \qquad (1)$$

A $\delta$ parameter of 0 represents a proton shared equally between two water molecules and is described as Zundel-like; in contrast, an increasing $\delta$ parameter represents a proton that is associated more with a single water molecule, which then forms a more Eigen-like complex (likely a distorted one that is not purely three-fold symmetric). However, while the $\delta$ parameter is a good first step in characterizing the hydrated proton complex, several authors[26, 40, 43] have questioned the validity of the $\delta$ parameter to always clearly distinguish the actual solvation structure of the hydrated excess proton. As an example, Daly et al.[43] stressed that since the distribution is non-bimodal it can be difficult at times to clearly distinguish the hydrated excess proton as being either distinctly Zundel or Eigen.

In a detailed simulation study, Swanson and Simons[26] first discussed the degree of information gained from the $\delta$ parameter. Specifically, by applying an *ab initio* energy decomposition analysis (EDA), they reported that the sum of the $\delta$ values within a distorted Eigen complex and Zundel complex can qualitatively track the degree of charge transfer within the first solvation shell of the hydrated excess proton (see Figures 6 and 7 in ref[26]). By tracking the $\delta$ value of the excess proton as a function of time during which the proton does not hop (i.e., the special pair "dance")[22], they quantified the $\delta$ value to be less than 0.1, which highlights the limits in the ability to apply $\delta$ parameter values toward characterizing the solvation structure of the hydrated excess proton from only a single snapshot. Swanson and Simons concluded that studying the actual dynamics of the hydrated excess proton can be more revealing for determining its solvation structure. With this definition in hand, therefore, Eigen-like dynamics can be more fully characterized by periods during which the identity of the special pair changes; conversely, Zundel-like dynamics can be characterized by proton hopping back-and-forth between two water molecules over an extended period of time with no special pair change. This broader definition is also supported elsewhere in the literature.[46]

By using various acid solution trajectories of hydrated excess protons with two different simulation methods – as well as instantaneous normal mode analysis – in the present paper we more clearly elucidate the dynamics of the hydrated excess proton which at the same time also provides a more revealing interpretation of the recent spectroscopic results. In particular, based on these trajectories, we are able to capture specific processes that give rise to the anisotropy decay. As detailed herein, we document the 2.5 ps timescale as corresponding to proton transport observed by recent nonlinear spectroscopy, while confirming the structure of the hydrated excess proton to be best described as a distorted Eigen cation. Additionally, we explore the instantaneous normal modes for the hydrated excess proton that reveal distorted-Eigen configurations at the 1750 cm$^{-1}$, band, while also detailing the limitations of the $\delta$ parameter to characterize these vibrational modes and the hydrated excess proton structure.

This paper is outlined as follows: In the next section, we describe the two simulation methods used to calculate the data, while also providing simulation details. In the next section, we then use radial distribution functions (RDFs) to show the solvation structure of the hydrated excess proton is best characterized as a distorted Eigen cation. We next provide our anisotropic dynamics data for the distorted-Eigen cation, as well as the time-evolution of normal modes related to the special pair dynamics. In the final section, we provide conclusions.

Methods

**Simulations**

Two different and complementary simulation methods were used to carry out this study: the Multistate Empirical Valence Bond (MS-EVB) Method[47-51] and Experiment Directed Simulation[52] for *Ab Initio* Molecular Dynamics (EDS-AIMD).[53] We briefly explain the methodology behind these approaches, but we direct the reader to the original literature for further details.

In the MS-EVB formalism, the hydrated excess proton charge defect is dynamically delocalized amongst multiple water molecules by creating a "wavefunction" that is a linear combination of different bonding topology states. In each diabatic state, a different water molecule is covalently bonded to the excess proton, such that

$$|\Psi\rangle = \sum_{i}^{N} c_i\, |i\rangle \quad (2)$$

Once the linear combination is constructed, the coefficients of the ground state function are determined through an eigenvalue problem using a quantum-like Hamiltonian, as follows:

$$\mathbf{Hc} = E_0 \mathbf{c} \quad (3)$$

where $E_0$ and the coefficients are functions of the instantaneous nuclear configuration of the system. In the Hamiltonian the diagonal elements are defined using molecular mechanics force field terms, augmented by a repulsive term to correct for an over-attraction of the water molecules to the core hydronium cation. At the same time, the off-diagonal elements are used to couple the different diabatic states and allow transitions between them to occur "on the fly" (and hence to model chemical bonding rearrangements) in the MS-EVB MD simulation.

For this paper, we used the MS-EVB 3.2 and (anharmonic MS-EVB) aMS-EVB 3.2 models.[51] In comparison to earlier MS-EVB models for the excess proton in water,[17, 48-49] an additional Lennard-Jones potential energy term was incorporated in the 3.2 version to better account for the fourth water pre-solvation around the hydronium core.[54] The anharmonic aMS-EVB 3.2 model also incorporates non-harmonic vibrations in solvating the water molecules. In previous work,[34, 45, 55] the MS-EVB 3.2 model was used to help interpret infrared spectroscopy data of hydrated excess protons in bulk HCl acid and isotopically substituted water solutions. The model also provided cluster configurations extracted from those systems to calculate instantaneous normal modes using electronic density functional theory (DFT) at the B3LYP functional level

To compare our results with another independent and complementary simulation method, we also utilized EDS-AIMD simulation.[53, 56] Such an approach corrects for the over-structuring and slow diffusion of water in BLYP/BLYP-D3-based AIMD simulations by including an additional minimal bias to the system Hamiltonian through a potential energy term (Eq. 4 below). In this study, two EDS-AIMD methods are used. The first corrects the system water oxygen-oxygen (Ow-Ow) radial distribution function (RDF) to better match experiment,[53] hence known as EDS-AIMD(OO), while the second corrects the system water oxygen-hydrogen (Ow-Hw) RDF to match MB-pol's classical water model for this RDF which is known to be highly accurate.[57-60] This method is known as EDS-AIMD(OH).[56] In either case, the additional EDS potential is given by

$$V(r_i) = \sum_{k=0}^{M} \frac{\alpha_k}{\hat{f}_k} \sum_{j \neq i}^{n_O} r_{ij}^k [1 - u(r_{ij} - r_0)] \quad (4)$$

which is a term added to the DFT potential in the AIMD. It includes the coupling constant ($\alpha_k$) and the experimental target values ($\hat{f}_k$). This term utilizes a mollified smoothing function to make the coordination number and its statistical moments continuous. The EDS-AIMD(OH) method utilized concepts from Li and Swanson[61] to bias the hydrogen bond in acidic solutions in a continuous manner and details will be published in a forth coming article from our group. The coupling constant in Eq. 4 for EDS-AIMD(OO) was parameterized so that the oxygen-oxygen coordination number and its first-through-third moments could accurately reproduce experimental results, while the coupling constant for EDS-AIMD(OH) was parameterized to match the oxygen-hydrogen coordination number and its second moment of MB-pol's O-H RDF.[57-60] By directly biasing the coordination number and its moments, other structural and dynamical properties were also seen to be improved without increased AIMD computational cost.[53, 56] Most importantly for the present study, the properties of the hydrated excess proton are also significantly improved through the EDS-AIMD approach, especially the ratio of the excess proton diffusion to water diffusion.

**Simulation Details**

In total, ten independent MS-EVB 3.2 and aMS-EVB 3.2 simulations of 1 HCl in 256 $H_2O$ with a box side length of 19.73 Å were first equilibrated in the constant NVT ensemble using a Nose-Hoover thermostat with a temperature set to 298 K and a time-constant of 50 fs using an in-house version of the LAMMPS simulation package.[62] Water molecules were modelled using SPC/Fw[63] and aSPC/Fw[50] for MS-EVB 3.2 and aMS-EVB 3.2, respectively. After a 1 ns non-reactive equilibration period, each simulation was equilibrated using our reactive molecular dynamics simulation code for 500 ps in the constant NVT ensemble, and all production runs were carried out in the constant NVE ensemble for 1 ns. Each MS-EVB simulations used a timestep of 0.5 fs with a long-range cutoff of 9.0 Å and an Ewald summation with an error of $10^{-5}$.

In total, three EDS-AIMD(OO) simulations of 1 HCl in 128 water molecules with a box side length of 15.72 Å and two EDS-AIMD(OH) simulation of an $H^+$ in 128 water molecules with a box side length of 15.64 Å were equilibrated in the constant NVT ensemble using a Nose-Hoover thermostat with a time-constant of 11.12 fs, followed by 80 ps and 200 ps in the constant NVE ensemble for EDS-AIMD(OO) and EDS-AIMD(OH), respectively. EDS parameters were identified from our previous work[53, 56] for the BLYP exchange-correlation functional[64-65] with the D3 Grimme dispersion interaction.[66-67] All EDS-AIMD simulations were carried out with Quickstep module in CP2K[68] and PLUMED[69-70] packages using Goedecker-Teter-Hutter (GTH) pseudopotentials[71] with a TZV2P basis set having a plane-wave cutoff of 400 Ry.

**Instantaneous Normal Mode Analysis**

Instantaneous normal mode (INM) calculations[72-73] were carried out to further compare structural vibrations and infrared absorption frequencies, with the goal of better clarifying the solvation structure of the hydrated excess proton. The INM calculations were derived as in a prior report,[34] but were augmented here by using both MS-EVB 3.2 and EDS-AIMD(OO) trajectories. The hydrated excess proton was identified as the proton in the hydronium cation with the lowest $\delta$ value; all water molecules within a 5 Å radius were included in the NMA calculation. Further NMA calculations restricted the number of water molecules to include only those in the second solvation shell, since this approach was shown to capture the extent of charge-delocalization of the positive charge defect in the excess proton complex.[26] All NMA calculations were calculated using the B3LYP exchange-correlation functional[74] with a 6-31 G(p,d) basis set in the Gaussian 09 software package.[75] It should also be noted that each NMA was carried out in the absence of energy minimization, and so all imaginary frequencies were discarded prior to analysis.

Results and Discussion

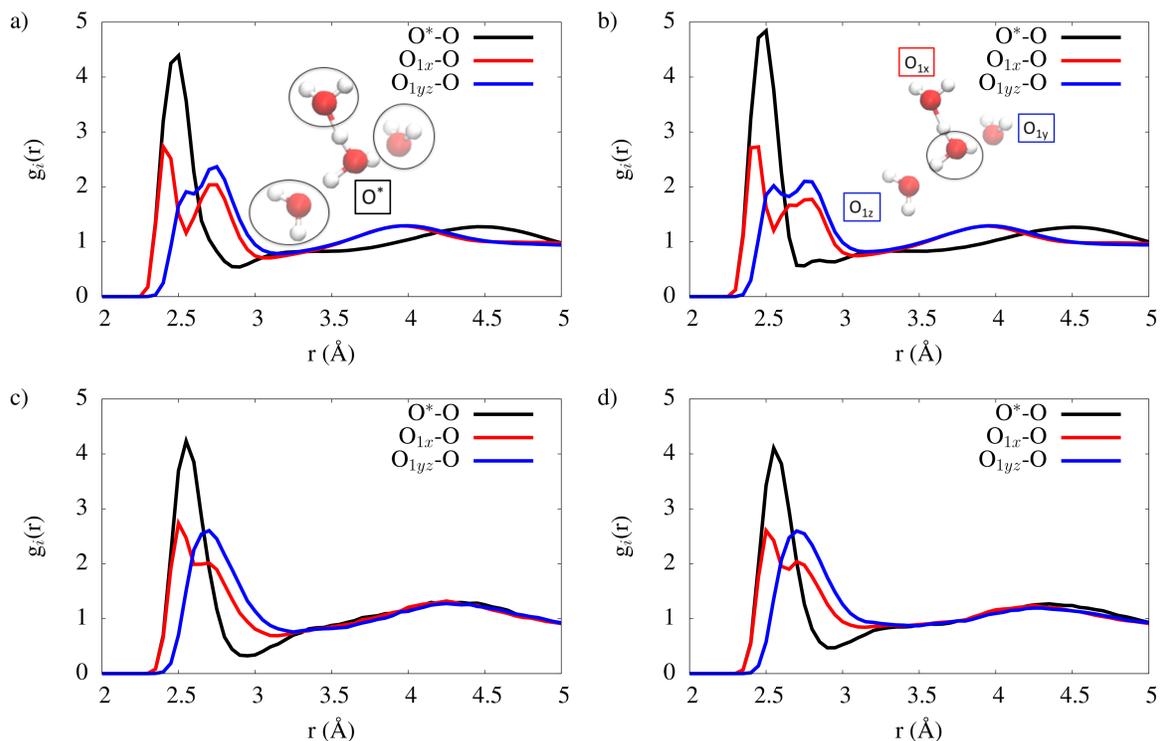

**Figure 1**. O-O radial distribution functions of the O*-Ow (black), $O_{1x}$-Ow (Red), and $O_{1yz}$-Ow (Blue) for (a) MS-EVB 3.2, (b) aMS-EVB 3.2 (c) EDS-AIMD(OO), and (d) EDS-AIMD(OH). O* is the oxygen with the most hydronium like character (most probable).

**Solvation Structure of Hydrated Excess Proton**

Figure 1 illustrates the O-O RDF for the various simulation methods. The O-O RDFs include all oxygen atoms in the system, those centered on the most probable hydronium-like (O*, black), the special pair oxygen ($O_{1x}$, red), and the remaining water molecules in the hydrated proton complex ($O_{1yz}$, blue). In these calculations, $O_{1x}$, $O_{1y}$ and $O_{1z}$ are defined as the neighboring water molecules with increasing $\delta$ value around the most probable hydronium.

In all simulation methods, the $O^*$-$O_w$ RDFs exhibit a unimodal peak centered at ~ 2.5 Å, which is attributed to three water molecules fluctuating around the hydronium care, as determined by integrating the first peak (the coordination number of ~3). However, the $O_{1x}$ RDFs show a prominent peak at a shorter distance, representing the interaction of the special pair $O_{1x}$ with the hydronium-like oxygen atom, while subsequent peaks correspond to the $O_{1x}$ interaction with second solvation shell of the $O^*$. This trend was also identified in the $O_{1yz}$ RDF, with the exception that the EDS-AIMD method is unimodal. As determined via EDS-AIMD, the $O^*$-$O_{1yz}$ and $O_{1yz}$-$O_w$ interactions were seen to be quite similar, which explains the unimodal peak in the $O_{1yz}$; this finding is further supported by the $O_{1yz}$ RDF having an integrated value close to 4.

We note that the solvation structures of the hydronium-like structure and the special pair oxygen are not identical, as indicated by different $O^*$ and $O_{1x}$ RDFs. In a symmetric Zundel cation, the excess proton would remain in the center of the two flanking water molecules, which would then make these flanking waters nearly identical. Even when the excess proton "rattling" is factored into this Zundel picture, the ensemble average of this rattling would still give rise to identical flanking water molecules and nearly identical solvation structures for the $O^*$ and $O_{1x}$, which is not observed in any of these simulation results. We also point out that recent studies that have proposed a symmetric Zundel cation as the dominate species in solution have distinguished Zundel and Eigen cation differently from the present work, with a criterion that is based on the $O_{1x}$ distance being less than or greater than 2.7 Å, respectively, for Zundel and Eigen. [40] This classification method is questionable, however, since we note that for the simulation methods reported here, all $O_{1x}$ distances were found to be less than 2.7 Å, so the criterion used on this other cited work would never identify a distorted Eigen cation as has been done here. A purely Zundel cation picture also is at odds with the difference in the in O* and $O_{1x}$ RDFs seen here from four different simulation approaches.

The calculated RDFs in this work indeed indicate a distinction between the hydronium-like structure having the excess proton (O*) and the special pair (O$_{1x}$), which is a clear characteristic of a distorted Eigen cation. (We note that the distorted Eigen cation has also been suggested as the most thermodynamically stable structure in recent AIMD studies which additionally used activated rate theory to characterize proton transfers in water.[76]) In a two-water shared proton configuration picture,[43] we see clear evidence of a single water molecule positioned closer to the hydronium-like cation, which is in agreement with the RDFs found in Figure 1. However, these various static snapshots ignore the important dynamics of the protonated complex. If the two-water picture also represented the actual dynamics of the system, the O*-O$_w$ RDF would match the O$_{1x}$ RDF, corresponding to a dominant first peak for the special pair and integrating to a coordination of a single water molecule. In contrast, the ensemble and time-averaged structure shows a single peak in the O*-Ow RDF and an average coordination of approximately three water molecules. Given these findings – coupled with the preference for the excess proton to associate with one water molecule at a given instant – the simulation evidence strongly suggests that a distorted Eigen best characterizes the hydrated excess proton. By virtue of the distorted Eigen picture, we can incorporate the special-pair (i.e., instantaneous two-water shared proton picture), while additionally accounting for the dynamics of the hydrated excess proton involving other water molecules in the larger Eigen cation complex. In fact, the picture of a distorted Eigen cation as best representing the hydrated excess proton structure first emerged from simulation more than twenty years ago.[17]

**Special-Pair Dance Revisited**

It has previously been shown that the hydrated excess proton is not a static complex, but is very dynamic one. In prior work,[22] it was revealed that the identity of the excess proton's special pair switches with other waters in the hydrated complex on a short timescale of tens of femtoseconds. Here we revisit this result using the EDS-AIMD(OH) hydrated excess proton model, which is the most recent model and one we consider to be the most accurate (including the use of the highly accurate MB-pol model[57-60] as the EDS reference).

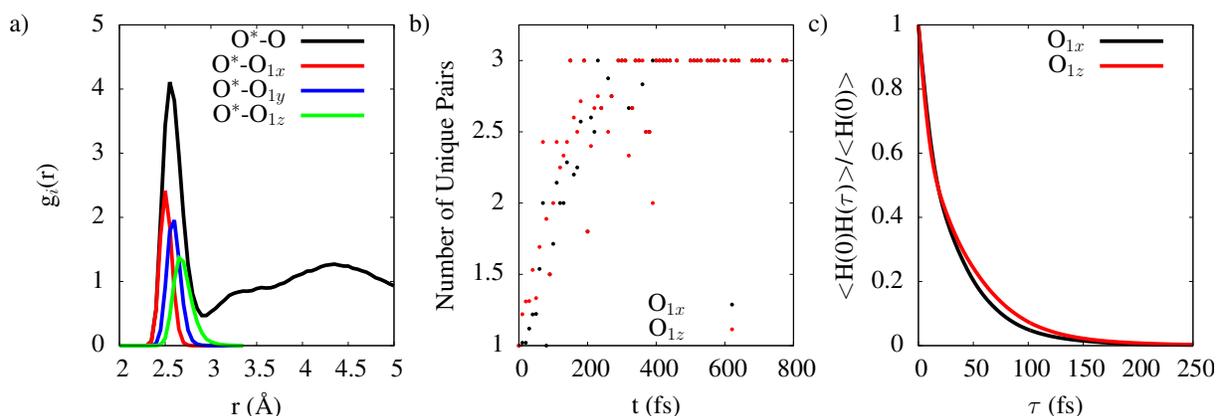

**Figure 2.** (a) EDS-AIMD(OH) O*-Ow RDF (black) decomposed into its three water molecules ($O_{1x}$ (red), $O_{1y}$ (blue), and $O_{1z}$ (green)) in the distorted Eigen cation based on increasing $\delta$ values. (b) The number of unique $O_{1x}$ (black) and $O_{1z}$ (red) pairs for periods where the proton does not hop. (c) Continuous correlation function of the $O_{1x}$ (black) and $O_{1z}$ (red).

In Figure 2a we show more detail of the O*-Ow RDF of the EDS-AIMD(OH) model from Fig. 1d. We further decompose the first peak into is special pair ($O_{1x}$) and corresponding water molecules ($O_{1y}$ and $O_{1z}$) in the distorted Eigen complex. We again defined these water molecules with increasing $\delta$ value. When examining Fig. 2a, three distinct peaks are found with varying distances from the most hydronium-like oxygen, which correspond with the closest water being the special pair. The overall statistically combined RDF peak (black line) is clearly unimodal. This result is another indication of the distorted Eigen cation as being the predominant hydrated proton species, where the three water molecules in the 1st shell are statistically unique when decomposed into three component peaks, and one water molecule on average is found to be closer to the hydronium core. However, this is an average over all configurations and the specific water molecules contributing to these three peaks vary with time.

We next investigate the dynamics of the closest and furthest water molecule in Figs. 2b and 2c. We begin by examining the number of unique $O_{1x}$ and $O_{1z}$ pairs for periods of time where the most hydronium-like oxygen does not change (Fig 2b). For short times, there is only one unique water molecule for the special-pair and the third furthest water molecule. When the proton resides on a single water molecule for longer times, we find 3 unique water molecules could be identified as the special-pair and the third furthest water molecule. These findings indicate that when the proton resides on a single water molecule for short times – such as the case for rattling events – only one water molecule is found to be the special-pair and another is found to at the furthest distance ($O_{1z}$). This is also the case when only considering snapshots of configurations. On the other hand, during longer periods when the excess proton remains on a single water molecule, the special-pair and corresponding $O_{1z}$ water molecules are dynamically switching in the distorted Eigen complex. These results point to the significance of accounting for all 3 water molecules in the proton complex instead of using static configurations such as is often the case for a two-water picture.

We additionally calculated the continuous correlation function for the special pair and the $O_{1z}$ water molecule according to the following equation

$$C(t) = \frac{\langle h(0)h(t) \rangle}{\langle h(0) \rangle} \tag{5}$$

where $h(t) = 1$ during segments of the trajectory where the $O_{1x}$ or $O_{1z}$ do not change and 0 for all other times. This is similar to the continuous correlation function used for the excess proton structure.[77] We find that the continuous correlation function for both the $O_{1x}$ and $O_{1z}$ are very similar. By integrating these correlation functions, we can gain an estimate of the lifetime of these species in the distorted Eigen cation, and we find a lifetime of 15.7 fs and 17.3 fs for $O_{1x}$ and $O_{1z}$, respectively. These again point to the dynamical nature of the surrounding water molecules in the distorted Eigen cation, with the special pair continuously evolving between the hydronium core and several (three) water molecules dynamically solvating it, with excess proton rattling events occurring between all of these three waters but at different times.

**O\*-Ow Anisotropy Decay**

Recent nonlinear infrared spectroscopy data from Carpenter et al. indicates that the reorientation timescale of the hydrated excess proton complex is about 2.5 ps,[38] which was confirmed by calculating the anisotropy decay via parallel and perpendicular pulses at 1740 - 1790 cm$^{-1}$. In this important new experimental work, the vibrational bending of the flanking water molecules in the special pair was assigned to 1750 cm$^{-1}$. These authors also de-emphasized certain phenomena – notably, the complete reorientation of the hydrated excess proton complex without proton transfer, rapid structural fluctuations, and energy and thermal transfer from the hydrated complex to surrounding aqueous environment – as possible structural reorientations that might explain the 2.5 ps timescale. Instead, they suggested that the 2.5 ps reorientation timescale corresponds to irreversible proton transfer. Additionally, they claimed that special pair dance could not play a role in their results, as one might have expected a change in the identity of the excess proton to reorient the transition dipole more rapidly than 2.5 ps.

Here, we seek to understand through simulation the structural reorientations of the hydrated excess proton complex that correspond to the observed experimental anisotropy timescales, with special emphasis on elucidating the nature of the special pair dynamics and irreversible proton transport. It is very common in MD simulations of water to calculate the anisotropy of the O-H stretch by using the unit vector along the O-H bond.[78] Given recent findings using a two-water or special pair approach, we decided to define the unit vector along the O*-Ow axis of the special pair. Anisotropy calculations were then determined using the second Legendre polynomial (Eq. 5) of the unit vector defining the special pair, such that

$$C_2(t) = \frac{\langle P_2(\hat{u}(t)\hat{u}(0))\rangle}{\langle P_2(\hat{u}(0)\hat{u}(0))\rangle} \quad (6)$$

Here, the $\langle ... \rangle$ denotes both an ensemble average and a time average of the unit-vector. In our molecular simulations, the special pair is defined as the O*-Ow that has the lowest $\delta$ parameter.

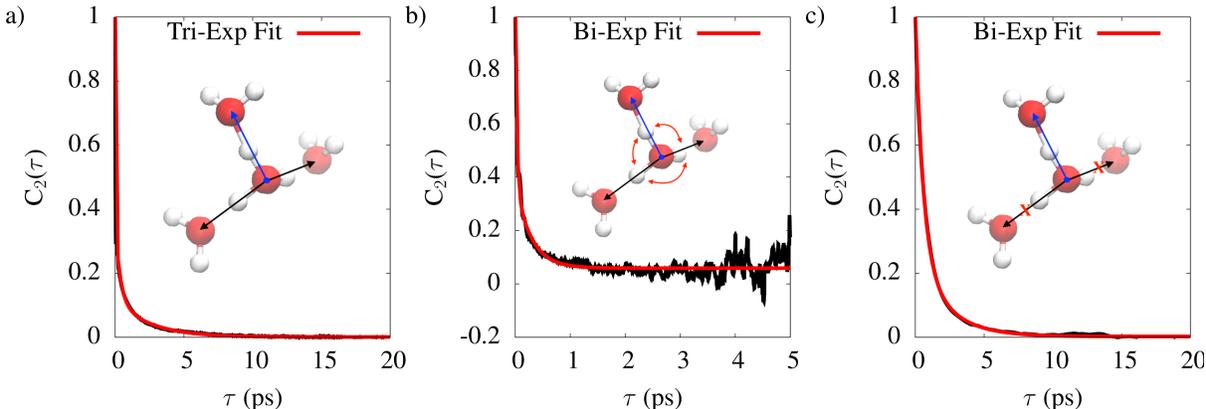

**Figure 3**. Anisotropy plots using Eq. 5 for the O*-Ow unit vector. The anisotropy plots are broken down based on (a) total anisotropy, (b) special pair "dance", and (c) long-lived special pair.

Figure 3a illustrates the anisotropy decay of the O*-Ow special pair from MS-EVB 3.2 simulations (corresponding O*-Ow anisotropy plots for aMS-EVB, EDS-AIMD(OO) and EDS-AIMD(OH) can be found in Section S1 of the Supporting Information, denoted later as SI). By fitting a triple exponential fit to total anisotropy decay (see Discussion in Section S2 of SI), we obtained time constants of 12 fs, 0.36 ps, and 2.47 ps, with corresponding exponential term amplitudes of 0.65, 0.23, and 0.12. Comparable values were obtained from the other simulation methods, as indicated in Table 1. Of particular importance is that these time constants and amplitude data can be replicated across various simulation methods, confirming that our findings are not unique to the MS-EVB simulations. It is also to be noted that one of the three processes described by the tri-exponential fit is very close to the 2.5 ps time constant determined experimentally.[38] To identify the precise structural phenomena corresponding to these time constants, we therefore removed specific structural dynamics from the trajectory input for the anisotropy decay calculations and refit that data to a bi-exponential function. Since we observed good agreement between the MS-EVB and EDS-AIMD simulations, we conducted subsequent calculations using only the more computationally efficient MS-EVB approach, as it can be run much longer and thus provide better statistics in comparison to the EDS-AIMD approach.

**Table 1.** Tri-exponential fits to the Total Anisotropy Decays (eq 5)

| System | $a_1$ | $\tau_1$ (fs) | $a_2$ | $\tau_2$ (ps) | $a_3$ | $\tau_3$ (ps) | C |
|---|---|---|---|---|---|---|---|
| MS-EVB | 0.65 | 12 | 0.23 | 0.36 | 0.12 | 2.47 | 0.00 |
| aMS-EVB | 0.66 | 10 | 0.23 | 0.32 | 0.11 | 2.27 | 0.00 |
| EDS-AIMD(OO) | 0.74 | 17 | 0.17 | 0.49 | 0.09 | 3.19 | -0.01 |
| EDS-AIMD(OH) | 0.74 | 17 | 0.19 | 0.38 | 0.08 | 2.57 | 0.00 |

First, we removed from the anisotropy calculation any contribution in the trajectory that was associated with proton transfer, and then decomposed the total anisotropy based on the special pair dance. As noted earlier, this is a process wherein the identity of the excess proton changes in a distorted Eigen cation, but with the absence of any change in the identity of the central hydronium-like core. We then parsed the trajectory into segments where the proton remained on a single water molecule, and next calculated the unit vector for the special pair. The anisotropy was then calculated for each segment and averaged over all segments. The special pair dance anisotropy calculation is shown in Figure 3b for MS-EVB 3.2; as indicated therein, we obtained time constants of 28 fs and 0.29 ps from a bi-exponential fit. Note that these bi-exponential constants are quite comparable to those obtained using the other simulation methods (Table 2). Previous MS-EVB simulations have shown that the identity of the special pair on tens of fs timescales, which is consistent with the 28 fs identified in this anisotropy calculation. This rapid timescale is too fast to be resolved by current experimental techniques; in fact, it is faster than the shortest pulse used experimentally.

**Table 2.** Bi-exponential fits to the Special pair Dance Anisotropy Calculations

| System | $a_1$ | $\tau_1$ (fs) | $a_2$ | $\tau_2$ (ps) | C |
|---|---|---|---|---|---|
| MS-EVB | 0.68 | 28 | 0.32 | 0.29 | 0.06 |
| aMS-EVB | 0.70 | 29 | 0.30 | 0.31 | 0.05 |
| EDS-AIMD(OO) | | | | | |
| EDS-AIMD(OH) | | | | | |

In a similar manner, we examined the slower time constant by removing the special pair dance from the total anisotropy calculations. Using this approach, we no longer define the special pair as the O*-Ow pair with the lowest $\delta$ value, but rather as the O-O vector between the hydronium oxygen and the water oxygen to which the excess proton hops. For example, if at timestep 0 the hydronium oxygen is molecule A, and at timestep $t$ the hydronium oxygen is molecule B, then the special pair unit vector from timestep 0 to timestep $t$ is defined as the O-O unit-vector between molecule A and B. The anisotropy calculation for MS-EVB 3.2 is found in Figure 3c. Table 3 provides amplitude and time constant data for the various methods. Note that for MS-EVB 3.2 we identified time constants of 0.56 ps and 2.17 ps, which are in close agreement with the intermediate and longer time constants listed in Table 1. The 2.17 ps result is also in good agreement with the 2.5 ps reported from the non-linear spectroscopy. It should also be pointed out that in all three tables there is seen an intermediate timescale of the 0.3 - 0.5 ps. We did not specifically analyze this motion but since it involves the decay of an angular correlation we presume it reflects the diffusive rotation of the overall Eigen complex.

Table 3. Bi-exponential fits to the Long-Lived Anisotropy Calculations

| System | $a_1$ | $\tau_1$ (ps) | $a_2$ | $\tau_2$ (ps) | C | $R^2$ |
|---|---|---|---|---|---|---|
| MS-EVB | 0.74 | 0.56 | 0.26 | 2.17 | 0.00 | 0.91 |
| aMS-EVB | 0.67 | 0.53 | 0.33 | 1.80 | 0.00 | 0.90 |
| EDS-AIMD(OO) | | | | | | |
| EDS-AIMD(OH) | | | | | | |

It should be appreciated that we were able to analyze these anisotropy plots while fully keeping the distorted-Eigen cation picture of the hydrated excess proton complex. By removing the special pair dance from the distorted-Eigen cation we were able to recover the long-lived time constant; analogously, by removing proton transport component we were able to retain the special pair dance. In addition to confirming the agreement between the time constants and amplitudes between the bi-exponential fits to the total tri-exponential fits, these findings indicate that (a) the fast time constant correlates to the special pair dance, and (b) the slow time constant corresponds to irreversible proton transfer. These hypotheses are supported by the good agreement of the timescales and physical processes obtained from recent 2D-IR experiments. However, all of this agreement is obtained without having to do away with the distorted Eigen cation picture.

**Instantaneous Normal Mode Analysis using 5 Å Radial Cutoff**

Our simulation methods show good agreement with the anisotropy timescales determined experimentally. However, the issue of the experimental evidence for the special pair dance prompted an investigation reported herein of the correspondence between the special pair dance and specific vibrational frequencies. With this process depicted as a series of dynamic switches between the hydrated excess proton special pair, we utilized configurations within a 100 fs timespan from the MS-EVB and EDS-AIMD(OO) results that correspond to the special pair dance. Instantaneous Normal Mode (INM) analysis was then applied to configurations separated by 1 fs, which differs from prior gas phase NMA calculations that pick randomly selected configurations; accordingly, our method enables us to observe the time-evolution of the normal modes.

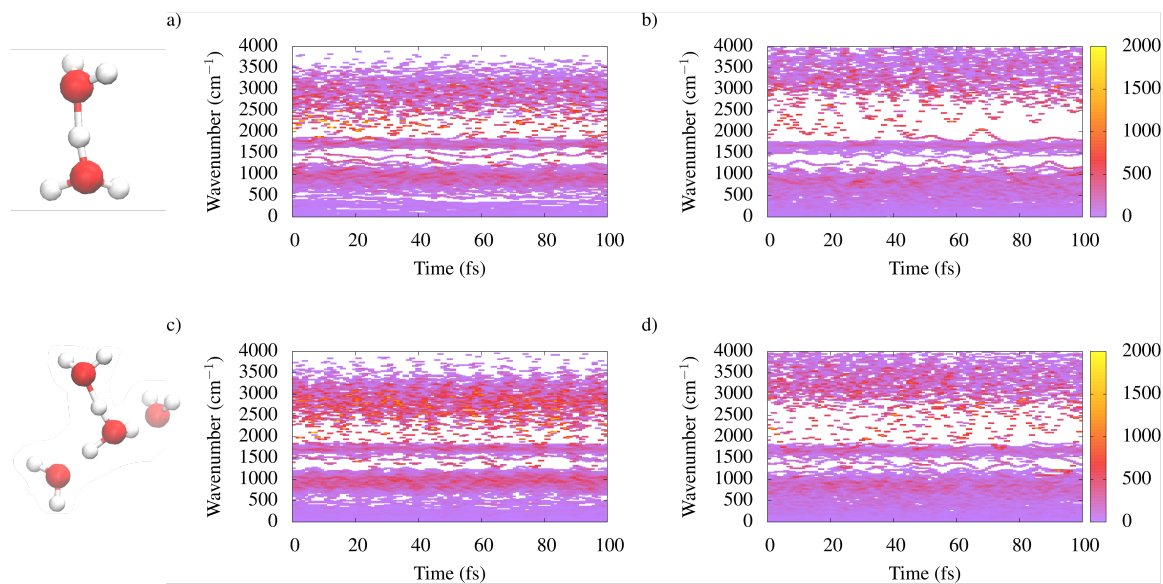

**Figure 4.** Spectral density as a function of time for Zundel (a & b) and Eigen (c & d) configurations using configurations taken from MS-EVB 3.2 (a & c) and EDS-AIMD (b & d). Zundel and Eigen are defined based on having 1 (Zundel) or 3 (Eigen) unique special pair water molecules during the 100 fs segment.

Two 100-fs segments obtained from MS-EVB 3.2 and EDS-AIMD simulations were selected where the hydronium-like molecular index remained unchanged. In the first segment, the identity of the special pair changed between the three water molecules; in contrast, the identity of the special pair remained unchanged in the second segment. We defined the first 100-fs segment as "Eigen", and the second as "Zundel". All waters and hydronium molecules whose oxygen atoms were found to reside within a 5 Å radial cutoff of the excess proton were included in the INM calculations. It should be noted that during the Eigen segment, the identity of the excess proton changed and was subsequently identified as the excess proton within the special pair. With the change in identity of the excess proton, the center of the 5 Å cutoff will also change; as such, some water molecules would be removed from the INM calculation and included in the configurations.

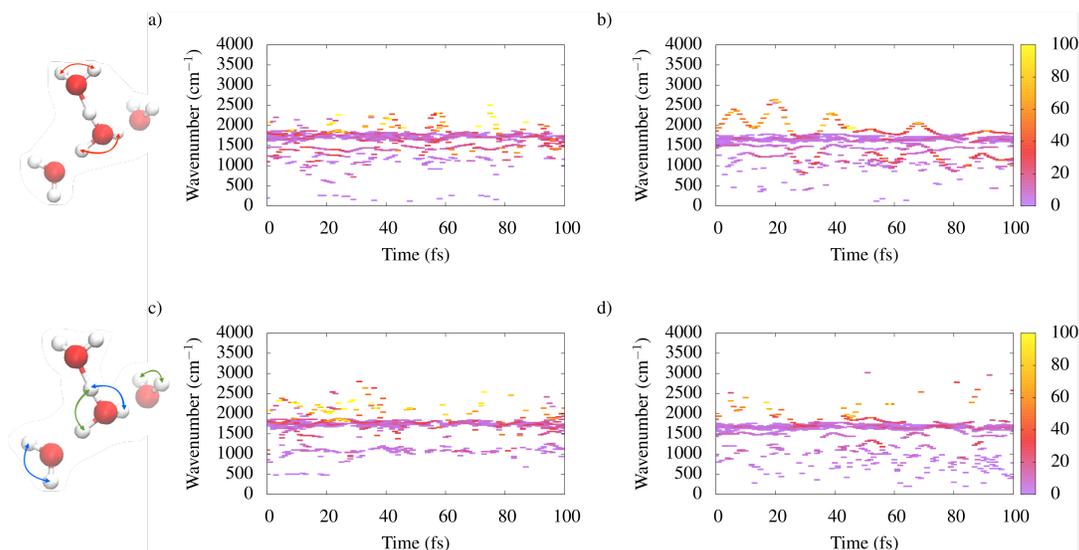

**Figure 5.** Spectral density as a function of time for H-O-H bends in Zundel configurations using data obtained from MS-EVB 3.2 (a & c) and EDS-AIMD (b & d) simulations. Panels a and b show the spectral density for H-O-H bends >5° in the special pair, while Panels c and d show the spectral density of H-O-H bends >5° in other O*-Ow pairs.

Figure 4 shows the spectral density of the normal modes as a function of time, with configurations pulled from MS-EVB 3.2 and EDS-AIMD(OO) trajectories. Upon first inspection we see many similarities in the spectral density plots, independent of the simulation method and whether the 100 fs segments were defined as Eigen or Zundel. The most striking difference is the mean distribution of the OH stretches obtained from MS-EVB and EDS-AIMD simulations. Of special interest is the frequency range around 1750 cm$^{-1}$, since this data corresponds to the special pair flanking water bend. As indicated in Figure 4, there is a persistent spectroscopic signature at this frequency for <u>both</u> Zundel and Eigen configurations, and it is not found only within a Zundel complex.

We next selected normal modes from these calculations that exhibited flanking water bending within the Zundel trajectory (Figure 5) and Eigen trajectory (Figure 6). If both flanking H-O-H angles in the special pair (Figure 5a,b and Figure 6the a,b) or other O*-Ow pairs in the first solvation shell (Fig, 5c,d and Figure 6c,d) changed by more than 5 degrees, normal modes were then pulled and their spectral densities were plotted. For each timestep, the spectral density was then divided by the number of selected normal modes within that timestep to help compare the time-evolution of the spectral density. In both the Zundel and Eigen trajectories, H-O-H bending could be found at 1750 cm$^{-1}$ that did not result from the special pair, but rather from other O*-Ow pairs in the distorted-Eigen cation. This finding confirms that the 1750 cm$^{-1}$ resonance is independent of the structure classification, which is in good agreement with prior work indicating that vibrations at 1750 cm$^{-1}$ can delocalize across 5-20 atoms.[34] This outcome may also explain why the experimentally obtained data could not confirm the special pair dance, in that many H-O-H bending vibrations can be excited instead of only the special pair itself.

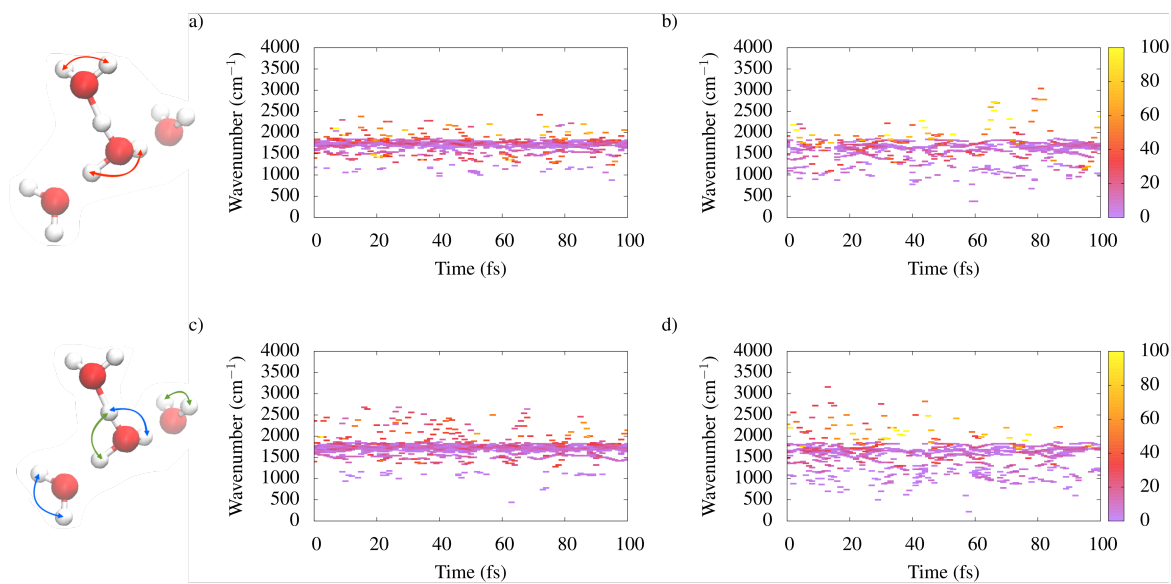

**Figure 6.** Spectral Density as a function of time for H-O-H bends in an Eigen configurations using configurations taken from MS-EVB 3.2 (a & c) and EDS-AIMD (b & d). Panels a and b show the spectral density for H-O-H bends >5° in the special pair, while Panels c and d show the spectral density of H-O-H bends >5° in other O*-Ow pairs.

**Normal Mode Analysis Using Second Solvation Shell**

Wave-like oscillations observed in some of the normal modes (specifically in Figure 5a,b and Figure 6a,b) could arise from an inconsistency in the number of water molecules in the normal mode calculations due to the identity change of the hydrated excess proton. Accordingly, we restricted the number of water molecules to include only those within the second solvation shell of the hydronium core as this was found to contain most of the charge transfer within the hydrated excess proton complex. Figure S4 of the SI illustrates the density of states and spectral density from these two kinds of INM calculations for the Eigen cation. Note the excellent agreement between the 5 Å cutoff and the second solvation method, as well as nearly identical spectral densities shown in Figures 7-9 below when we restrict the calculation to only the second solvation shell (Figure S5). Additionally, we obtained nearly identical wave patterns for these instantaneous normal modes (see Figures 5 and 6), which indicates that the oscillations in these normal modes are not due water molecules moving in and out of the NMA calculation.

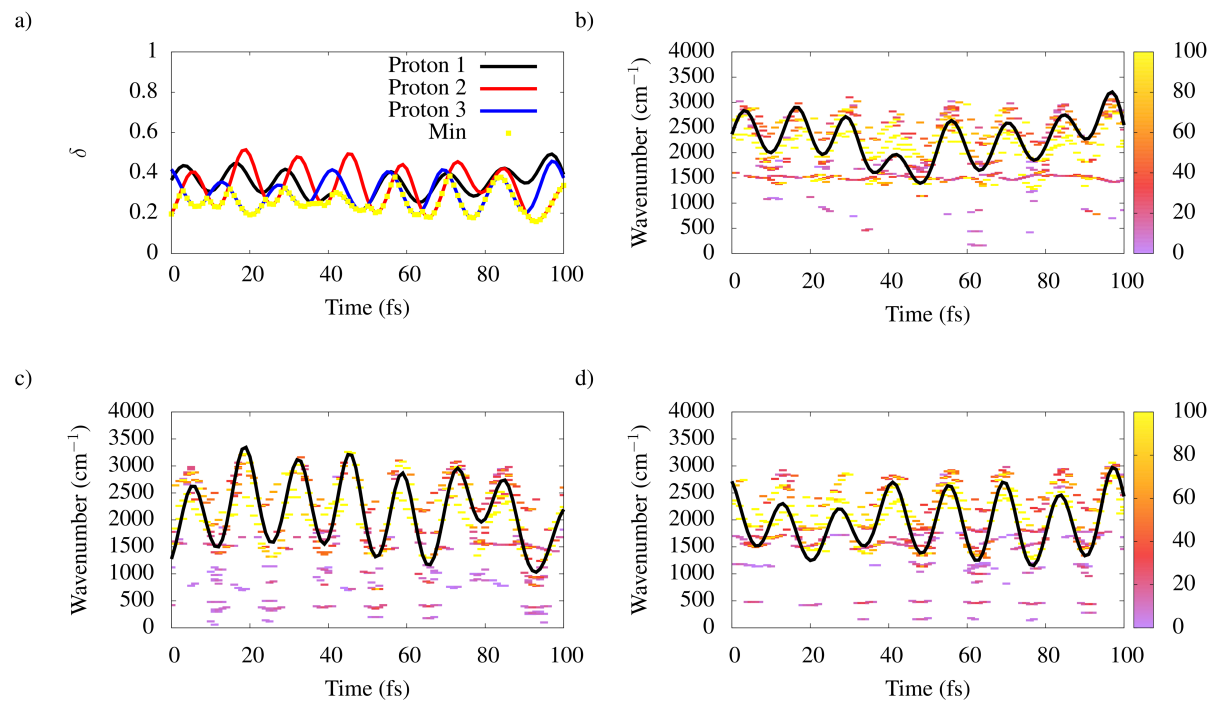

**Figure 7.** Spectral density and $\delta$ value for the hydronium-like protons in an Eigen configuration of an MS-EVB 3.2 simulation. Panel a shows the $\delta$ values for the three protons. The proton with the lowest $\delta$ value (i.e., the excess proton) is highlighted in yellow. Panels b-d shows the spectral density for protons 1, 2, and 3, respectively. Normal modes are selected if each proton's O*-$H_H$ stretches more than 0.2 Å. In black we overlay each proton's $\delta$ value multiplied by 6,500 to fit the figure.

Oscillatory patterns were also identified by Swanson and Simons in the charge transfer of the hydrated excess proton; moreover, they observed that the $\delta$ parameter qualitatively agreed with the trend in charge transfer (see Figure 4 in Ref. [26]). They also proposed defining Eigen and Zundel cations based on the dynamics of the excess proton rather than the $\delta$ parameter. In time segments more characteristic of a distorted Eigen cation, they observed instances during which the $\delta$ parameter became less than 0.1 Å. Such findings would typically indicate a Zundel cation; however, when placed in the context of the dynamics of the complex, it would instead show a distorted Eigen since the special pair is rotating among the three waters. They further proposed that Zundel cations could be characterized by an excess proton rattling between two water molecules. In following text, we therefore address how the $\delta$ value can track the charge transfer in distorted Eigen cation with and without a $\delta$ value less than 0.1 Å, and a Zundel cation described by a proton rattling between two waters.

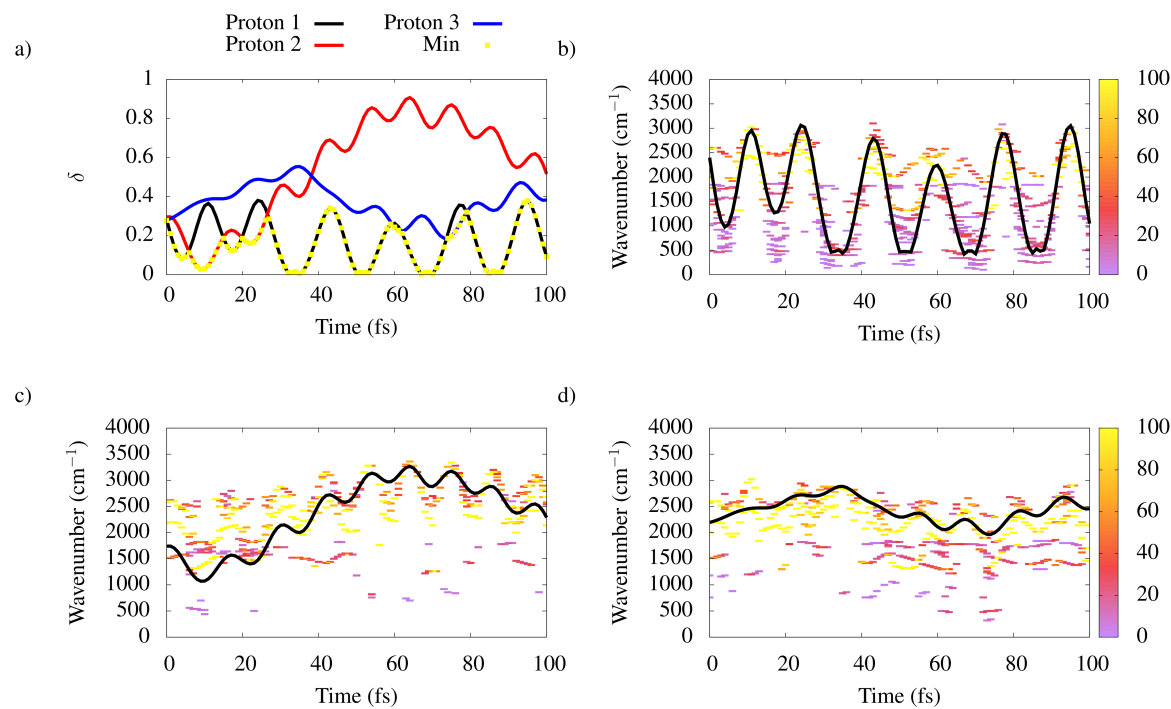

**Figure 8.** Spectral density and $\delta$ value for the hydronium's protons in a distorted Eigen configuration of MS-EVB 3.2. Figure a shows the $\delta$ values for the three protons. The proton with the lowest $\delta$ value (i.e. the excess proton) is highlighted in yellow. Figure b-d shows the spectral density for protons 1, 2, and 3, respectively, with each proton's $\delta$ value scaled to fit the figure in black. Proton 1's $\delta$ value was scaled by $\delta*7{,}000 + 400$, Proton 2's $\delta$ value was scaled by $\delta*2{,}500 + 1000$, and Proton 3's $\delta$ value was scaled by $\delta*2{,}500 + 1500$. Normal modes are selected if each proton's O*-$H_H$ stretches more than 0.2 Å.

We begin by examining the INM of the *Eigen* configurations of Fig. 4 and 6 which had the identity of the excess proton change within the *Eigen* complex without any change in the central hydronium core. In Figure 7a we show the $\delta$ value for the three protons in the complex. We assigned yellow squares to indicate the proton with the lowest $\delta$ value, which is used to distinguish the excess proton. In this segment, we find that the $\delta$ values of all three protons are very comparable with very similar mean values. We specifically point out that this is highly characteristic of the special pair dance, where the identity of the excess proton alternates between the other protons in the distorted Eigen complex. Figure 7 b-d illustrates the spectral density for each hydrogen in cases when the $O_H$-$H_H$ distance of the hydronium molecule changes by more than 0.2 Å, and, we superimposed the $\delta$ value for each proton over its spectral density features and scaled them to fit the plot. In this particular distorted Eigen segment, the $\delta$ value of the excess proton never goes below 0.1 Å, but nevertheless switches with the other protons in the complex. However, we find noticeable agreement between the shape of the normal mode oscillations and the $\delta$ parameter alluding to the charge transfer nature of each proton in the normal modes, and these specific normal modes are close to the bulk-like water stretches around 2500 cm$^{-1}$.

We did the same analysis on another Eigen configuration, but this time we examined a 100 fs which incorporates the special-pair dance that has a proton with a $\delta$ value less than 0.1 Å (Figure 8a). In this configuration, proton 1 begins as the excess proton and is the excess proton for the majority of the 100 fs, with proton 2 competing as the excess proton at the beginning of the segment, and proton 3 competing as the excess proton at the end of the segment. Additionally, the $\delta$ value of proton 1 is found in the region typically identified as Zundel, although, in light of the special-pair dance, it is best described as a distorted Eigen cation. We show the corresponding normal modes for this configuration in Fig. 8b-d. For proton 1 (Fig. 8b) we see spectroscopic signatures that are much more red-shifted in comparison to the other two protons in the distorted Eigen cation. Comparably, when proton 2 competes with proton 1 as the excess proton, we find some normal modes around 1750 cm$^{-1}$, yet as the simulation continues, its $\delta$ value increases and its corresponding normal modes are blue shifted to larger frequencies around 2750 cm$^{-1}$, which are much closer to bulk-like water stretches. This trend is found in the reverse order for proton 3, where at the beginning of the 100 fs segment, it has the largest $\delta$ value and therefore has the smallest charge transfer to its adjacent water molecule. However, at around 60 fs, proton 3 competes with Proton 1 as the excess proton and normal modes are slightly red shifted.

Figure 9 depicts the instantaneous normal modes for protons in a Zundel complex from a 100 fs time segment using the definition of Swanson and Simons. Water molecules found within first two solvation shells of each oxygen in the Zundel cation were selected and used for normal mode analysis. Proton 1 is the excess proton, while protons 2-5 represent the protons in the flanking waters. In all cases, each proton's $\delta$ value follows the shift in their corresponding normal mode. As indicated in Figure 9c, proton 1 has spectroscopic signatures in the 1000 – 2000 cm$^{-1}$ range that contains the PTM (1200 cm$^{-1}$) and the flanking water bend (1750 cm$^{-1}$) when the O-H bond in the normal mode changes by 0.2 Å. Conversely, proton 2 within the Zundel complex exhibits a larger frequency signature than the excess proton (~ 2900 cm$^{-1}$), and additionally, it displays a red shift in its normal mode frequency when its $\delta$ value closely approximates the $\delta$ value of proton 1 (~80-100 fs in Figure 9d), showing a greater degree of charge transfer and characteristics similar to those of the excess proton.

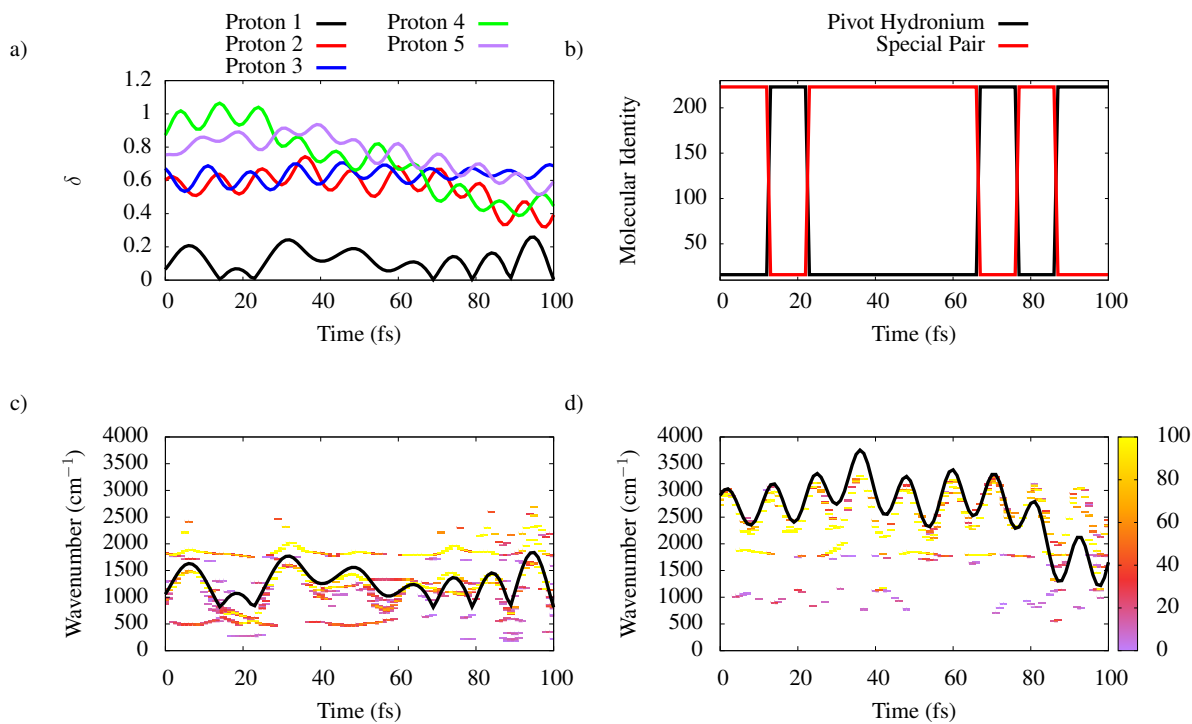

**Figure 9.** Spectral density and $\delta$ value for the protons in a Zundel configuration of MS-EVB 3.2. Panel a shows the $\delta$ values for the excess proton (Proton 1) and the 4 flanking protons (Protons 2-5). Panel b show the dynamics of the hydronium-like molecular identity ("pivot" hydronium is the water with the most hydronium like character at that instant). Panels c and d shows the spectral density for the excess proton (Protons 1) and proton 2 overlaid with their $\delta$ value scaled by $\delta*4{,}000 + 1500$ and $\delta*6{,}000\text{-}700$, respectively. Normal modes are selected if each proton's $O^*\text{-}H_H$ stretches more than 0.2 Å.

While the quantitative results described herein do not imply a direct correlation between structural geometries and instantaneous normal modes, they do further validate the work of Swanson and Simons. Specifically, due to the strong connection between the $\delta$ parameter and charge transfer within the hydrated excess proton, we propose that the oscillations in the instantaneous normal modes are driven by the charge transfer between each proton and its neighboring water molecule. Moreover, our findings indicate that the instantaneous normal modes correlate to the identity of the excess proton; specifically, each proton's normal modes red-shift as their $\delta$ parameter decreases (and respective charge transfer increases) as shown in the two Eigen configurations and Zundel configurations. Thus, it appears that the $\delta$ parameter can assist in determining the single proton that is the "excess proton"; however, determining the excess proton without the context of the overall dynamics of the hydrated proton complex can be misleading as is clearly seen in the Eigen configuration with $\delta$ values less than 0.1 Å..

Conclusions

In this work, we have utilized MS-EVB and EDS-AIMD simulations to further explore the structure of the hydrated excess proton, in the light of recent spectroscopic experiments. As such, we have obtained anisotropy decay data that closely match those obtained via nonlinear spectroscopy. By decomposing the anisotropy based on structural phenomena, we were able to identify anisotropy timescales that give rise to the special pair dance and the long-time decay of irreversible proton transfer. These processes associated with these timescales agree with previous theoretical studies of the hydrated excess proton,[21-22] while replicating the 2.5 ps anisotropy decay timescales achieved experimentally.[38] Perhaps most importantly, all of these results were obtained from several simulation methods that show the distorted Eigen cation, continuously undergoing a special pair dance, as the dominant "core" hydrated proton structure in dilute acid solution.

By examining the spectroscopic signatures of the various protons within limiting structures of Eigen and Zundel cations from the simulation data and INM analysis, we observed that the $\delta$ parameter tracks the shift in the normal mode of each proton, which has been shown to also characterize the charge transfer within the hydrated excess proton complex. Additionally, we noted that the excess proton defined by the lowest $\delta$ value around the hydronium molecule has a distinct spectroscopic signature in comparison to other protons in the complex. However, it must be noted that the identity of the excess proton is constantly changing prior to the occurrence of an irreversible proton transfer event (i.e., the special pair dance).[22] We conclude, therefore, that the $\delta$ parameter can help to characterize the excess proton structure, but at the same time we emphasize that selecting static structures from MD simulations and inferring spectroscopic information from them can be problematic without examining the actual dynamics. Moreover, an even simpler interpretation based on a one-dimensional effective potential picture (e.g., a symmetric or distorted Zundel cation as inferred from such a simple symmetric or asymmetric double well picture) is not able to capture the actual hydrated proton structure and dynamics.

Experimental anisotropy decay data for the flanking water bend show that increasing the chloride concentration decreases the proton transfer rate, while Arrhenius plots of the anisotropy timescales indicate that increasing the chloride concentration lowers the activation energy of irreversible proton transport.[39] Transition state theory arguments were then used to suggest that additional chloride ions, rather than excess protons, create entropic barriers to irreversible proton transport. Similarly, recent theoretical work[76] has also shown a concentration dependence of the O*-Cl ion pairing and proton transfer rate using transition state theory and Marcus theory, while also finding the Eigen cation as the thermodynamic stable state and the Zundel cation as a reaction intermediate. However, this latter work did not address the existence of an entropy barrier to the proton transport. By contrast, our very recent analysis[41] has provided an atomistic-level explanation for these phenomena by examining the influence of temperature and concentration on excess proton structure and transport in concentrated HCl solutions. This effort has also further reinforced the conclusions reached in the present study.

Author Information

Corresponding Author

*E-mail gavoth@uchicago.edu

Notes:

The authors declare no competing financial interest.

**Acknowledgments**

This research was supported by the U.S. Department of Energy, Office of Basic Energy Sciences, Separation Science Program of the Division of Chemical Sciences, Geosciences, and Biosciences under Award Number DE-SC0018648. The computational resources for this research were in part provided by the University of Chicago Research Computing Center (RCC), and in part from a computational grant from the U.S. Department of Defense (DOD) High Performance Computing Modernization Program at the Navy DOD Supercomputing Resource Centers.

**For Table of Contents Only**

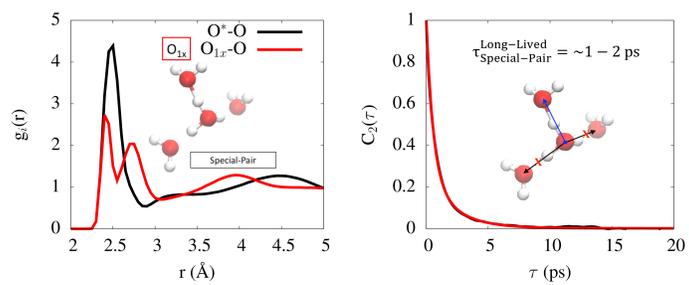

# Supporting Information for:

# Understanding the Essential Nature of the Hydrated Excess Proton Through Simulation and Interpretation of Recent Spectroscopic Experiments

Paul B. Calio, Chenghan Li, and Gregory A. Voth

Department of Chemistry, Chicago Center for Theoretical Chemistry, James Franck Institute, and Institute for Biophysical Dynamics, The University of Chicago, 5735 South Ellis Avenue, Chicago, Illinois 60637, United States

**S1. Anisotropy decays for aMS-EVB 3.2, and EDS-AIMD**

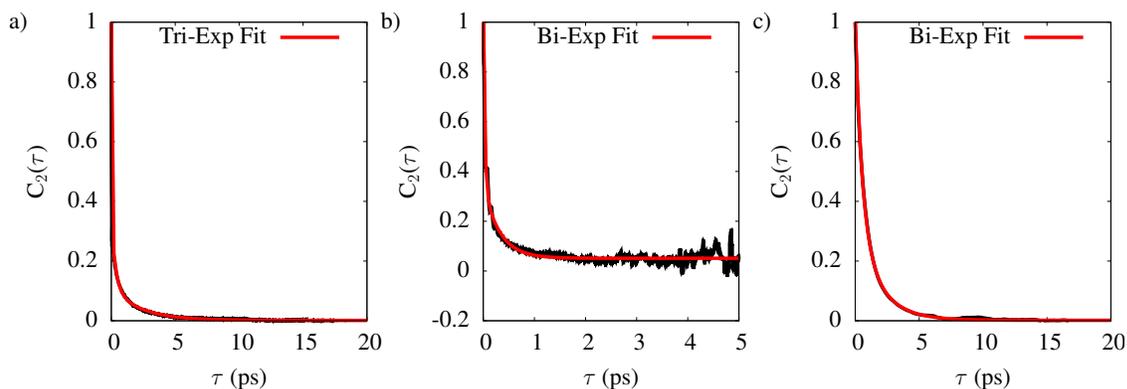

**Figure S1**. Anisotropy plots for the O*-Ow unit vector in aMS-EVB 3.2. The anisotropy plots are broken down based on (a) total anisotropy, (b) special-pair dance, and (c) long-lived special-pair.

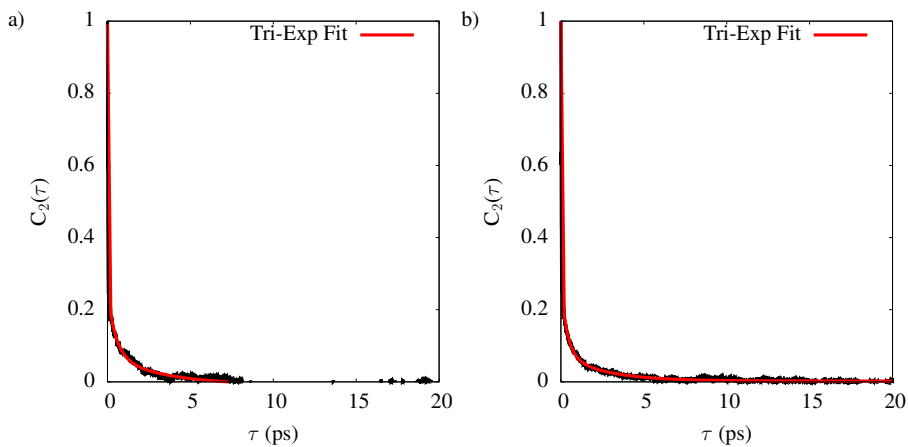

**Figure S2**. Total anisotropy plots for the O*-Ow unit vector in (a) EDS-AIMD(OO) and (b) EDS-AIMD(OH). Limited statistics prevent us from calculating the EDS-AIMD anisotropy of special-pair dance and long-lived special-pair.

## S2. Justification for Anisotropy Fitting Method

$$f(x) = a_1 \cdot \exp(-x/\tau_1) + a_2 \cdot \exp(-x/\tau_2) + a_3 \cdot \exp(-x/\tau_3) + C$$
$$\text{where in the fitting procedure, } a_3 = (1 - a_1 - a_2)$$

**Total Anisotropy**

Experimental anisotropy curves with fit using a bi-exponential fit. However, we found using tri-exponential fits best fit the curve as going from bi-exponential (Table S1) to tri-exponential (Table S2) increase the $R^2$ value on average increases by 0.01.

Table S1: Bi-exponential Fit for Total Anisotropy

| System | $a_1$ | $\tau_1$ (fs) | $a_2$ | $\tau_2$ (ps) | C | $R^2$ |
|---|---|---|---|---|---|---|
| MS-EVB | 0.742 | 17.5 | 0.258 | 1.143 | 0.005 | 0.988 |
| aMS-EVB | 0.751 | 14.2 | 0.249 | 1.021 | 0.004 | 0.986 |
| EDS-AIMD(OO) | 0.806 | 21.5 | 0.194 | 1.402 | -0.005 | 0.975 |
| EDS-AIMD(OH) | 0.807 | 21.6 | 0.193 | 1.048 | 0.0054 | 0.983 |

Table S2: Tri-exponential Fit for Total Anisotropy

| System | $a_1$ | $\tau_1$ (fs) | $a_2$ | $\tau_2$ (ps) | $a_3$ | $\tau_3$ (ps) | C | $R^2$ |
|---|---|---|---|---|---|---|---|---|
| MS-EVB | 0.649 | 12.3 | 0.234 | 0.363 | 0.117 | 2.472 | 0.001 | 0.998 |
| aMS-EVB | 0.661 | 10.1 | 0.229 | 0.324 | 0.110 | 2.267 | 0.001 | 0.997 |
| EDS-AIMD(OO) | 0.742 | 17.4 | 0.172 | 0.488 | 0.085 | 3.194 | -0.009 | 0.984 |
| EDS-AIMD(OH) | 0.736 | 17.2 | 0.188 | 0.377 | 0.076 | 2.572 | 0.002 | 0.992 |

**Special-Pair Dance (No Hopping)**

We fit the special-pair dance data to a bi-exponential curve as we expect the long-time, proton transfer process to be disregarded. When going from bi-exponential (Table S3) to tri-exponential (Table S4), the $R^2$ value on average increases by 0.001. Additionally, MS-EVB can be an approximate bi-exponential fit since the third time constants is 0.0, and aMS-EVB does not have any time constants longer than 1 ps, meaning the long-time scale is essentially removed from the anisotropy decay.

Table S3: Bi-exponential Fit for Special-Pair Dance

| System | $a_1$ | $\tau_1$ (fs) | $a_2$ | $\tau_2$ (ps) | C | $R^2$ |
|---|---|---|---|---|---|---|
| MS-EVB | 0.681 | 28.1 | 0.319 | 0.287 | 0.059 | 0.934 |
| aMS-EVB | 0.703 | 28.6 | 0.297 | 0.313 | 0.051 | 0.964 |
| EDS-AIMD(OO) | | | | | | |
| EDS-AIMD(OH) | | | | | | |

Table S4: Tri-exponential Fit for Special-Pair Dance

| System | $a_1$ | $\tau_1$ (fs) | $a_2$ | $\tau_2$ (ps) | $a_3$ | $\tau_3$ (ps) | C | $R^2$ |
|---|---|---|---|---|---|---|---|---|
| MS-EVB | 0.642 | 34.5 | 0.289 | 0.311 | 0.069 | 0.000 | 0.059 | 0.934 |
| aMS-EVB | 0.437 | 15.4 | 0.409 | 0.091 | 0.154 | 0.516 | 0.049 | 0.967 |
| EDS-AIMD(OO) | | | | | | | | |
| EDS-AIMD(OH) | | | | | | | | |

**Long-Lived Special Pair, No "Dance"**

We fit the long-lived, special-pair data to a bi-exponential curve as we expect the short-time, special-pair dance to be disregarded When going from bi-exponential (Table S5) to tri-exponential (Table S6), the $R^2$ value either remained constant or decreased when going from the bi-exponential fit to the tri-exponential fit. This seems to suggest that a bi-exponential fit can capture the proper physics of the system. Additionally, MS-EVB is essential a bi-exponential fit since the first time constant is close to zero.

Table S5: Bi-exponential Fit for Long-Lived Special-Pair

| System | $a_1$ | $\tau_1$ (ps) | $a_2$ | $\tau_2$ (ps) | C | $R^2$ |
|---|---|---|---|---|---|---|
| MS-EVB | 0.737 | 0.557 | 0.263 | 2.169 | 0.003 | 0.907 |
| aMS-EVB | 0.672 | 0.533 | 0.328 | 1.795 | 0.002 | 0.902 |
| EDS-AIMD(OO) | | | | | | |
| EDS-AIMD(OH) | | | | | | |

Table S6: Tri-exponential Fit for Long-Lived Special-Pair

| System | $a_1$ | $\tau_1$ (fs) | $a_2$ | $\tau_2$ (ps) | $a_3$ | $\tau_3$ | C | $R^2$ |
|---|---|---|---|---|---|---|---|---|
| MS-EVB | 0.052 | 0.4 | 0.786 | 0.682 | 0.162 | 2.889 | 0.001 | 0.907 |
| aMS-EVB | 0.043 | 4.0 | 0.885 | 0.955 | 0.072 | 0.001 | 0.006 | 0.900 |
| EDS-AIMD(OO) | | | | | | | | |
| EDS-AIMD(OH) | | | | | | | | |

**S3 – DOS and Spectral Density comparison of 5A and 2nd Shell cut-off**

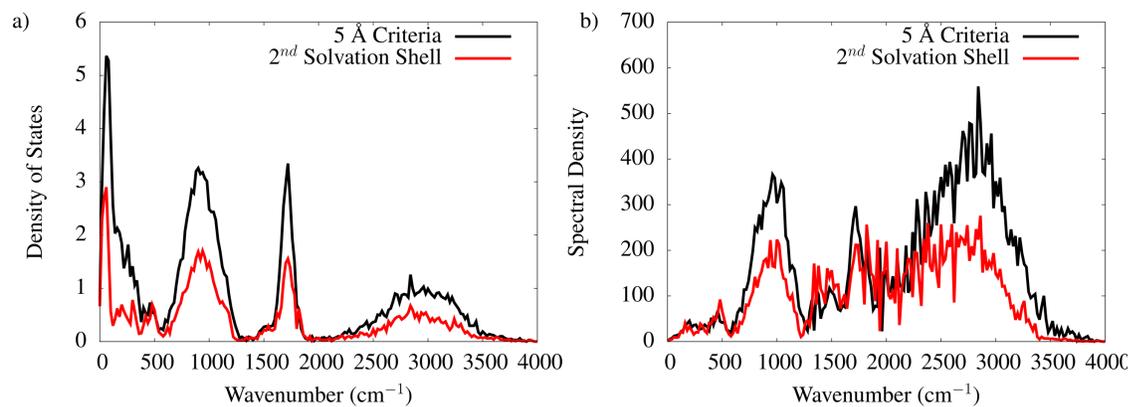

**Figure S3.** Density of States (a) and spectral density (b) from the INM analysis using the 5 Å cutoff and 2nd solvation shell cutoff.

## S4 – Spectral Density vs. time for 2nd solvation shell.

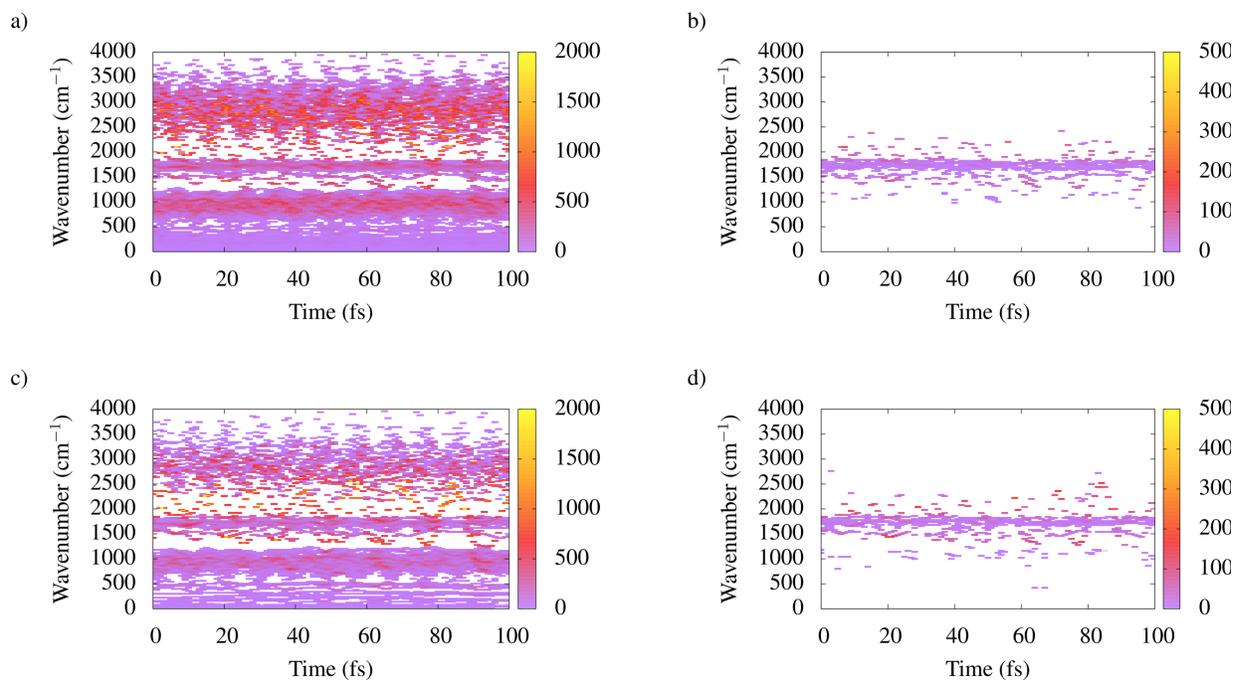

**Figure S4.** Spectral Density as a function of time for Eigen trajectories taken from MS-EVB 3.2 using the 5 Å cutoff (a & b) and 2nd solvation shell (c & d). Eigen trajectory is defined based on have 3 unique special-pair water molecules during the 100 fs segment. In Fig. (a) and (c) we show the total spectral density, while in Fig (b) and (d) we show the spectral density for H-O-H bends >5° in the special-pair